\documentclass[12pt]{article}
\usepackage{latexsym}
\usepackage{feynmp}

\newbox\SlashedBox  
\def\fs#1{\setbox\SlashedBox=\hbox{#1} 
\hbox to 0pt{\hbox to 1\wd\SlashedBox{\hfil/\hfil}\hss}{#1}} 
\def\hboxtosizeof#1#2{\setbox\SlashedBox=\hbox{#1} 
\hbox to 1\wd\SlashedBox{#2}} 

\def\ms#1{\setbox\SlashedBox=\hbox{$#1$}
\hbox to 0pt{\hbox to 1\wd\SlashedBox{\hfil/\hfil}\hss}#1}

\newcommand{\tr}{{\rm tr}}
\newcommand{\ie}{{\em i.e.~}}
\newcommand{\eg}{{\em e.g.~}}

\newcommand{\be}{\begin{equation}}
\newcommand{\ee}{\end{equation}}
\newcommand{\ba}{\begin{eqnarray}}
\newcommand{\ea}{\end{eqnarray}}

\begin{document}

\vspace*{-0.5cm}    
    
\begin{flushright}
ROM2F/99/31
\end{flushright}

\vspace{0.5cm}

\begin{center}

{\LARGE {\bf Non-renormalisation of extremal correlators 
in ${\cal N}$=4 SYM theory\rule{0pt}{25pt} }} \\
\vspace{1cm} {\large Massimo Bianchi and Stefano Kovacs} \\ 
\vspace{0.6cm} 
{\large {\it Dipartimento di Fisica, \ Universit{\`a} di Roma \  
``Tor Vergata''}} \\  {\large {\it I.N.F.N.\ -\ Sezione di Roma \ 
``Tor Vergata''}} \\ {\large {\it Via della Ricerca  Scientifica, 1}} 
\\ {\large {\it 00173 \ Roma, \ ITALY}} 

\end{center}

\vspace{0.5cm}

\begin{abstract}
 
We show that extremal correlators of chiral primary operators in
${\cal N}$=4 supersymmetric Yang--Mills theory with $SU(N)$ gauge 
group are neither renormalised at first ($g^2$) order in perturbation 
theory nor receive contribution from any instanton sector at leading 
order in the semiclassical expansion. This lends support to the strongest 
version of a new prediction recently put forward on the 
basis of the AdS/SCFT correspondence. 

\end{abstract}

\vspace{0.5cm}

\section{Introduction}

A new prediction of the correspondence between type IIB superstring
on $AdS_{5}\times S^{5}$ and ${\cal N}$=4 
supersymmetric Yang--Mills (SYM) theory with $SU(N)$ gauge 
group~\cite{jm,agmoo} is that a certain class of
$n$-point correlation functions, represented by ``extremal correlators"
of chiral primary operators (CPO's), satisfy some non-renormalisation 
theorem. In \cite{extremal}, D'Hoker, Freedman, Mathur, Matusis and 
Rastelli have argued that type IIB supergravity requires
the above $n$-point functions to decompose into products of $n-1$ 
free-field two-point functions. Moreover the overall coefficient was 
argued not to be renormalised with respect to its free field value.
Something similar is known to happen for two- and three-point functions 
of CPO's~\cite{lmrs,dfs,ehssw} as well as for the chiral~\cite{fmmr} 
and Weyl~\cite{sken} anomalies\footnote{The $1/N^{2}$ corrections to the 
chiral anomaly are reproducible in terms of one-loop corrections to the
lowest order supergravity approximation due to the absence in the 
bulk of the would-be singleton fields~\cite{bilchu}.}.

The extremal correlation functions we consider are of the form
\begin{equation}
    G_{\rm ext}(x,x_{1},\ldots,x_{n}) = \langle 
    {\cal Q}^{(\ell)}(x) {\cal Q}^{(\ell_{1})}(x_{1}) \ldots 
    {\cal Q}^{(\ell_{n})}(x_{n}) \rangle \; , 
    \label{qextrcorr}
\end{equation}
with $\ell=\ell_{1}+\ell_{2}+\ldots+\ell_{n}$.
The operators ${\cal Q}^{(\ell)}$ in (\ref{qextrcorr}) are CPO's, \ie
scalar composite operators of protected dimension $\Delta=\ell$ 
belonging to the representation with Dynkin labels $[0,\ell,0]$ of the 
$SU(4)$ R-symmetry group. They are lowest components of short $SU(2,2|4)$ 
supermultiplets. Other shortening conditions are possible, see 
\cite{dp,fz} for a detailed discussion.

For single-trace CPO's one has
\begin{equation}
    {\cal Q}^{i_{1}i_{2}\ldots i_{\ell}}(x) = 
    \sum_{{\rm perms}~\sigma} \tr \left[ \varphi^{\sigma(i_{1})} 
    \varphi^{\sigma(i_{2})}\ldots 
    \varphi^{\sigma(i_{\ell})} - ~{\rm flavour~contractions} \right]  \; .
    \label{defq}
\end{equation}
For multi-trace CPO's one has an obvious generalisation of 
(\ref{defq}).
The correlator (\ref{qextrcorr}) is ``extremal'' in that there is only 
one $SU(4)$ invariant contraction of the (implicit) ``flavour'' indices, 
\ie there is only one $SU(4)$ singlet in the 
product of the representations with Dynkin labels $[0,\ell,0]$, 
$[0,\ell_{1},0]$, \ldots, $[0,\ell_{n},0]$. 

It is the purpose of this letter to show that extremal correlators
of CPO's are neither renormalised at first order in perturbation 
theory\footnote{Dan Freedman has informed us that a similar result has 
been independently obtained by Witold Skiba.} 
nor receive contribution from any instanton sector at leading order in the
semiclassical expansion. 

The non-renormalisation properties displayed by extremal 
correlators of CPO's at
first order in perturbation theory suggest that the strong version of
the argument proposed in \cite{extremal} on the basis of the AdS/SCFT
correspondence should be valid. One might then expect that
any extremal correlators, either involving single-trace operators or
multi-trace operators, should be independent of the coupling constant 
$g$, hence tree-level exact, for any finite $N$.
If this were the case no higher derivative term in the type IIB 
superstring effective action should be capable of giving a 
non-vanishing amplitude of this kind. $SL(2,Z)$ invariance of 
type IIB superstring requires that string loop corrections to
higher derivative terms be accompanied by non-perturbative 
D-instanton corrections. It is by now widely appreciated that 
the counterpart of type IIB D-instantons in the AdS/SCFT correspondence
are SYM instantons \cite{bg,bgkr,dhkmv}. These are responsible for
interesting $U(1)_{B}$ violating processes such as a 16-dilatino
amplitude, that has been used for a quantitative test of the AdS/SCFT 
correspondence \cite{bgkr,dhkmv}, as well as for some $U(1)_{B}$ preserving 
processes such as the higher derivative corrections to the four 
stress-tensor/graviton amplitude. Motivated by these considerations we 
have extended our analysis to the non-perturbative level. 

The plan of the letter is as follows.
After briefly reviewing the description of ${\cal N}$=4 supersymmetric 
Yang--Mills theory in terms of ${\cal N}$=1 superfields, we demonstrate
the vanishing of the lowest perturbative correction to the extremal 
correlators. We then pass to describe the fermion zero-mode counting in 
instanton backgrounds and show that non-perturbative corrections are absent 
as well. Finally we comment on the bearing and extension of our results in 
view of the AdS/SCFT correspondence.

We would like to stress that the results we are 
going to show are valid both for single- and for multi-trace operators 
in the relevant $SU(4)$ representation
and for any number of colours $N$, suggesting the validity of the
non-renormalisation theorem well beyond the reach of the lowest 
supergravity approximation, valid at large $N$ and large 't Hooft 
coupling. We suspect, but we do not explicitly show, 
that the result should hold for extremal correlators in ${\cal N}$=4 
SYM theories with other gauge groups. 

\section{${\cal N}$=4 SYM theory in the ${\cal N}$=1 formulation}
\label{n1form}

The field content of ${\cal N}$=4 SYM \cite{n4sym} is realised 
combining one 
${\cal N}$=1 vector superfield, $V$, with three ${\cal N}$=1 chiral 
superfields, $\Phi^{I}$ ($I$=1,2,3), all in the adjoint representation 
of the gauge group. The six real scalars, $\varphi^{i}$ 
($i$=1,2,\ldots,6), of the theory are assembled 
into three complex fields, namely 
\begin{equation}
    \phi^{I} = \frac{1}{\sqrt{2}} \left( \varphi^{I}+i\varphi^{I+3} 
    \right) \: , \qquad \phi^{\dagger}_{I} = \frac{1}{\sqrt{2}} \left( 
    \varphi^{I}-i\varphi^{I+3} \right) \; ,
    \label{defphi}
\end{equation}
that are scalar components of the superfields $\Phi^{I}$. Three 
of the Weyl fermions are the spinors of the chiral multiplets, denoted 
by $\lambda^{I}$, and the fourth spinor, $\lambda = \lambda^{0}$, together 
with the vector, $A_{\mu}$, form the vector multiplet. 
In this formulation only a $SU(3)\times U(1)$ subgroup of the original 
$SU(4)$ R-symmetry group is manifest, with $\Phi^{I}$ and 
$\Phi^{\dagger}_{I}$ transforming in the representations {\bf 3} and 
${\overline {\bf 3}}$ of $SU(3)$ respectively, while $V$ is a singlet. 

The action in the ${\cal N}$=1 superfield formulation reads
\begin{eqnarray}
    S & = & \frac{1}{l_{{\bf r}}g^{2}}\, \tr \left\{ \int d^{4}x \, 
    \left[ \left( \int d^{4}\theta \, \Phi^{\dagger}_{I}
    e^{{V}}\Phi^{I} \right) + \left( 
    \int d^{2}\theta \: \frac{1}{16} W^{\alpha}W_{\alpha} + {\rm h.c.}
    \right) \right. \right. + \nonumber \\
    & + & \left. \left. \!\!\frac{i\sqrt{2}}{3!} \left( \int d^{2}\theta \,  
    \varepsilon_{IJK} \Phi^{I}[\Phi^{J},\Phi^{K}] + \int d^{2}{\overline 
    \theta} \, \varepsilon^{IJK}
    \Phi^{\dagger}_{I}[\Phi^{\dagger}_{J},\Phi^{\dagger}_{K}] 
    \rule{0pt}{18pt} \right) \right] \right\} \; ,
    \label{n1superfield}
\end{eqnarray}
where $l_{{\bf r}}$ denotes the Dynkin index of the representation. 
In what follows we will mostly restrict our attention to the case of 
an $SU(N)$ gauge group. All the (super)fields belong to the 
adjoint representation, {\it viz.} 
\begin{equation}
    V(x,\theta,{\overline \theta}) = 
    V^{a}(x,\theta,{\overline \theta})T_{a} \: , \quad
    \Phi^{I}(x,\theta,{\overline \theta}) = 
    \Phi^{aI}(x,\theta,{\overline \theta})T_{a} \: , \quad 
    \Phi^{\dagger}_{I}(x,\theta,{\overline \theta}) = 
    \Phi^{a\dagger}_{I}(x,\theta,{\overline \theta})T_{a} \; .
    \label{defrepr}
\end{equation}
In equation (\ref{n1superfield}) the standard dependence on the coupling 
constant $g$ is recovered by substituting $V \rightarrow 2gV$, 
$\Phi^{I} \rightarrow g\Phi^{I}$ and $\Phi^{\dagger}_{I} \rightarrow 
g\Phi^{\dagger}_{I}$. Notice that the full action also contains the ghosts 
that are not displayed here since they do not contribute to the Green 
functions that we will consider at the order at which we compute 
them. 

Expanding the exponential $e^{2gV}$ in (\ref{n1superfield}) gives 
\begin{eqnarray}
    S &=& \int d^{4}x\:d^{4}\theta \, \left\{\rule{0pt}{18pt} 
    V^{a} \left[ - \Box +(1-\xi) (P_{1}+P_{2}) \Box \right] V_{a} 
    +\Phi^{a\dagger}_{I}\Phi^{I}_{a} + ig f_{abc}{\Phi^{\dagger}}^{a}_{I}
    V^{b}\Phi^{Ic} + \ldots \right. + \nonumber \\ 
    && \left. - \frac{\sqrt{2}}{3!} g f^{abc} \left[ \varepsilon_{IJK}
    \Phi_{a}^{I} \Phi_{b}^{J} \Phi_{c}^{K} \delta({\overline \theta}) + 
    \varepsilon^{IJK} \Phi^{\dagger}_{aI}\Phi^{\dagger}_{bJ}
    \Phi^{\dagger}_{cK}\delta(\theta) \right] + \ldots \right\} \; ,
\label{actionsuper}	
\end{eqnarray}
where $f_{abc}$ are the $SU(N)$ structure constants and terms that are 
not relevant for first order perturbative 
calculations have been neglected. The projection operators $P_{1}$ 
and $P_{2}$ are defined as
\begin{equation}
    P_{1}=\frac{1}{16}\frac{{\cal D}^{2}{\overline {\cal D}}^{2}}{\Box} 
    \:, \qquad
    P_{2}=\frac{1}{16}\frac{{\overline {\cal D}}^{2}{\cal D}^{2}}{\Box} \; ,
    \label{projectors}
\end{equation}
where ${\cal D}$ and ${\overline {\cal D}}$ are the standard 
super-covariant derivatives~\cite{wb}. 
In (\ref{actionsuper}) a gauge fixing term ($\xi = 1/\alpha$ is a 
gauge parameter) 
\begin{equation}
    S_{{\rm g.f.}} = \frac{1}{l_{{\bf r}}g^{2}} \tr \int d^{4}x \int 
    d^{2}\theta\,d^{2}{\overline\theta} \left[ -\frac{\xi}{32}
    \left({\cal D}^{2}V\right)\left({\overline {\cal D}}^{2}V\right) 
    \right] 
    \label{gaugefix}
\end{equation}
has been introduced as well.

We choose to work in components and without fixing the Wess-Zumino 
gauge. By expanding chiral and vector superfields according to 
\begin{eqnarray}
    && \hspace*{-0.9cm}
    \Phi^{I}(x,\theta,{\overline\theta}) = \phi^{I}(x)+\sqrt{2}\theta 
    \lambda^{I}(x) + \theta \theta F^{I}(x) + i\theta \sigma^{\mu}
    {\overline \theta}\partial_{\mu} \phi^{I}(x) + \frac{1}{\sqrt{2}}
    \theta \theta {\overline \theta} {\overline \sigma}^{\mu}
    \partial_{\mu}\lambda^{I}(x) + \nonumber \\
    && \hspace*{0.9cm}
    +\frac{1}{4} \theta \theta {\overline \theta} {\overline \theta} 
    \Box \phi^{I}(x) \label{chiral} \; , \\
    && \hspace*{-0.9cm}
    V(x,\theta,{\overline \theta}) = C(x) + i \theta \chi(x) - i 
    {\overline \theta} {\overline \chi}(x) + \frac{i}{\sqrt{2}}\theta 
    \theta S(x) -\frac{i}{\sqrt{2}} {\overline \theta}{\overline \theta} 
    S^{\dagger}(x) - \theta \sigma^{\mu}{\overline \theta} A_{\mu}(x) +
    \nonumber \\ 
    && \hspace*{-0.9cm}
    +i\theta \theta {\overline \theta} \left[ {\overline \lambda}(x) + 
    \frac{i}{2}{\overline \sigma}^{\mu}\partial_{\mu}\chi(x) \right] 
    -i{\overline \theta}{\overline \theta} \theta \left[ 
    \lambda(x) +\frac{i}{2}\sigma^{\mu}\partial_{\mu}{\overline \chi}(x) 
    \right] +\frac{1}{2} \theta \theta {\overline \theta}{\overline \theta} 
    \left[ D(x) + \frac{1}{2} \Box C(x) \right]  
\label{vector}
\end{eqnarray}
in the action, one obtains the equivalent component-field formulation.
In the Fermi--Feynman gauge, $\alpha=1$, the kinetic part of the action 
in components reads \cite{kov}
\begin{eqnarray}
    && S_{0} = \int d^{4}x \left[ 
    \phi^{a\dagger}_{I}\Box \phi_{a}^{I} - 
    {\overline \lambda}_{I}^{a}{\overline \sigma}^{\mu} 
    (\partial_{\mu}\lambda^{I}_{a}) + F^{a\dagger}_{I}F^{I}_{a} - 
    S^{\dagger}_{a}\Box S^{a} + \frac{1}{2} A^{a}_{\mu}\Box A_{a}^{\mu} + 
    \right. \nonumber \\ 
    && -\frac{1}{2} \left(C^{a}\Box^{2}C_{a} + C^{a}\Box D_{a} + 
    D^{a}\Box C_{a} \right) +\frac{1}{2} \Big( 
    \chi^{a}\Box\sigma^{\mu}\partial_{\mu}{\overline\chi}_{a} +
    {\overline\chi}^{a}\Box{\overline\sigma}^{\mu}\partial_{\mu}
    \chi_{a} + \nonumber \\
    && \left. - \chi^{a} \Box \lambda_{a} - \lambda^{a}\Box 
    \chi_{a} - {\overline\chi}^{a}\Box{\overline\lambda}_{a} -
    {\overline\lambda}^{a}\Box{\overline\chi}_{a} \Big) 
    \rule{0pt}{16pt} \right] \; . 
    \label{sfree}
\end{eqnarray}

The trilinear couplings, that are needed to compute the 
first order, \ie $O(g^{2})$, perturbative corrections to  
$n$-point functions of scalar composite operators, such as the 
extremal correlators of CPO's, can be obtained 
from the action (\ref{actionsuper}) by an analogous expansion. The 
result is 
\cite{kov} 
\begin{eqnarray}
     && S_{{\rm int}} = \int d^{4}x \left\{ igf_{abc} \left[ \frac{1}{2}
     \phi_{I}^{a\dagger}D^{b}\phi^{cI} + \frac{1}{2} 
     \phi_{I}^{a\dagger}(\Box C^{b})\phi^{cI} - \frac{1}{2} 
     (\partial_{\mu}\phi_{I}^{a\dagger})C^{b}(\partial^{\mu}\phi^{cI}) 
     + \right. \right. \nonumber \\
     && +\frac{i}{2} \left( \phi_{I}^{a\dagger}A_{\mu}^{b}(
     \partial^{\mu} \phi^{cI}) - (\partial^{\mu}\phi_{I}^{a\dagger}) 
     A_{\mu}^{b}\phi^{cI} \right) + \frac{i}{\sqrt{2}}\left( 
     \phi_{I}^{a\dagger}S^{b\dagger}F^{cI}-F_{I}^{a\dagger}S^{b}
     \phi^{cI} \right) + \nonumber \\ 
     && - F^{a\dagger}_{I}C^{b}F^{cI} + \frac{i}{\sqrt{2}} \left( 
     {\overline \lambda}^{a}_{I}{\overline \lambda}^{b}\phi^{cI} - 
     \phi_{I}^{a\dagger}\lambda^{b}\lambda^{cI} \right) + \frac{i}{\sqrt{2}}
     \left( F^{a\dagger}_{I}\chi^{b}\lambda^{cI} - {\overline \lambda}^{a}_{I} 
     {\overline \chi}^{b}F^{cI} \right) + \nonumber \\
     && \frac{i}{\sqrt{2}} \left( \phi^{a\dagger}(\partial_{\mu}
     {\overline \chi}^{b})\sigma^{\mu}\lambda^{cI} + {\overline \lambda}^{a}_{I}
     {\overline \sigma}^{\mu}(\partial_{\mu}\chi^{b}) \phi^{cI} \right) +
     \frac{1}{2} \left( C^{b}{\overline \lambda}^{a}_{I}{\overline \sigma}^{\mu}
     (\partial_{\mu}\lambda^{cI}) - C^{b}(\partial_{\mu}
     {\overline \lambda}^{a}_{I}){\overline \sigma}^{\mu} \lambda^{cI} 
     \right)  + \nonumber \\ 
     && \left. -\frac{i}{2} {\overline \lambda}^{a}_{I}
     {\overline \sigma}^{\mu}\lambda^{cI}A_{\mu}^{b} \right] 
     - \frac{\sqrt{2}}{3!}gf_{abc} \left[ \rule{0pt}{16pt} 
     \varepsilon^{IJK} \left( 
     \phi^{a\dagger}_{I} \phi^{b\dagger}_{J}F^{c\dagger}_{K} - 
     \phi_{I}^{a\dagger}{\overline\lambda}^{b}_{J}{\overline\lambda}^{c}_{K} 
     \right) + \varepsilon_{IJK} \left( \phi^{aI}\phi^{bJ} F^{cK} + 
     \right. \right. \nonumber \\
     && \left. \left. \left. - \phi^{aI} \lambda^{bJ}\lambda^{cK} \right) 
     \rule{0pt}{14pt} \right] \right\} \; . 
     \label{strilinear}
\end{eqnarray}

\section{Perturbative non-renormalisation}
\label{pert}

In the following we will concentrate on  extremal 
correlators of the specific form 
\begin{equation}
    G(x,x_{1},\ldots,x_{n}) = \langle 
    \tr\Big[\big(\phi^{1}\big)^{\ell}(x)\Big] 
    \tr\Big[\big(\phi^{\dagger}_{1}\big)^{\ell_{1}}(x_{1})\Big] \ldots 
    \tr\Big[\big(\phi^{\dagger}_{1}\big)^{\ell_{n}}(x_{n})\Big] \rangle 
    \; . \label{extrcorr}
\end{equation}
Up to an overall 
non-vanishing Clebsch--Gordan coefficient, computing correlators of this kind  
is equivalent to computing the generic correlator (\ref{qextrcorr}). 

The tree-level contribution to (\ref{extrcorr}) corresponds to a 
diagram with $\ell$ lines exiting from the point $x$, which form $n$ 
different ``rainbows'' connecting $x$ to the points $x_{i}$, the $i$th 
rainbow containing $\ell_{i}$ lines 

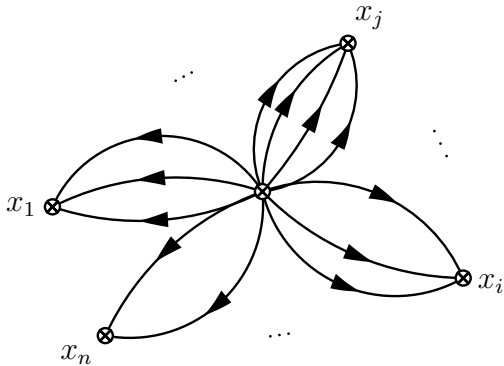
\vspace*{1cm}
\begin{center}
\begin{fmffile}{ext0}
 \begin{fmfgraph*}(160,110) 
  \fmfsurroundn{v}{32}
  \fmf{phantom,right=0.6,tag=1}{c,v17}
  \fmf{phantom,right=0.2,tag=2}{c,v17}
  \fmf{phantom,left=0.2,tag=3}{c,v17}
  \fmf{phantom,left=0.5,tag=4}{c,v7}
  \fmf{phantom,left=0.27,tag=12}{c,v7}
  \fmf{phantom,right=0.1,tag=5}{c,v7}
  \fmf{phantom,right=0.5,tag=6}{c,v7}
  \fmf{phantom,left=0.4,tag=7}{c,v31}
  \fmf{phantom,right=0.2,tag=8}{c,v31}
  \fmf{phantom,right=0.5,tag=9}{c,v31}
  \fmf{phantom,right=0.2,tag=10}{c,v21}
  \fmf{phantom,left=0.5,tag=11}{c,v21}
  \fmf{phantom,tag=12}{v9,v25}
  \fmffreeze
  \fmf{phantom,left=0.3,tag=13}{v17,v7}
  \fmf{phantom,left=0.3,tag=14}{v7,v31}
  \fmf{phantom,left=0.15,tag=15}{v31,v21}
  \fmfipath{p[]}
  \fmfiset{p1}{vpath1(__c,__v17)}
  \fmfiset{p2}{vpath2(__c,__v17)}
  \fmfiset{p3}{vpath3(__c,__v17)}
  \fmfiset{p4}{vpath4(__c,__v7)}
  \fmfiset{p5}{vpath5(__c,__v7)}
  \fmfiset{p6}{vpath6(__c,__v7)}
  \fmfiset{p7}{vpath7(__c,__v31)}
  \fmfiset{p8}{vpath8(__c,__v31)}
  \fmfiset{p9}{vpath9(__c,__v31)}
  \fmfiset{p10}{vpath10(__c,__v21)}
  \fmfiset{p11}{vpath11(__c,__v21)}
  \fmfiset{p12}{vpath12(__c,__v7)}
  \fmfiset{p13}{vpath13(__v17,__v7)}
  \fmfiset{p14}{vpath14(__v7,__v31)}
  \fmfiset{p15}{vpath15(__v31,__v21)}
  \fmfi{fermion}{subpath (0,length(p1)) of p1}
  \fmfi{fermion}{subpath (0,length(p2)) of p2}
  \fmfi{fermion}{subpath (0,length(p3)) of p3}
  \fmfi{fermion}{subpath (0,length(p4)) of p4}
  \fmfi{fermion}{subpath (0,length(p5)) of p5}
  \fmfi{fermion}{subpath (0,length(p6)) of p6}
  \fmfi{fermion}{subpath (0,length(p7)) of p7}
  \fmfi{fermion}{subpath (0,length(p8)) of p8}
  \fmfi{fermion}{subpath (0,length(p9)) of p9}
  \fmfi{fermion}{subpath (0,length(p10)) of p10}
  \fmfi{fermion}{subpath (0,length(p11)) of p11}
  \fmfi{fermion}{subpath (0,length(p12)) of p12}
  \fmfiv{d.sh=circle,d.fill=empty,d.si=2mm}{point length(0) of p1}
  \fmfiv{d.sh=cross,d.si=2mm}{point length(0) of p1}
  \fmfiv{d.sh=circle,d.fill=empty,d.si=2mm}{point length(p1) of p1}
  \fmfiv{d.sh=cross,d.si=2mm,la=$x_{1}$}{point length(p1) of p1}
  \fmfiv{d.sh=circle,d.fill=empty,d.si=2mm}{point length(p4) of p4}
  \fmfiv{d.sh=cross,d.si=2mm,la=$x_{j}$}{point length(p4) of p4}
  \fmfiv{d.sh=circle,d.fill=empty,d.si=2mm}{point length(p7) of p7}
  \fmfiv{d.sh=cross,d.si=2mm,la=$x_{i}$}{point length(p7) of p7}
  \fmfiv{d.sh=circle,d.fill=empty,d.si=2mm}{point length(10) of p10}
  \fmfiv{d.sh=cross,d.si=2mm,la=$x_{n}$}{point length(10) of p10}
  \fmfi{dots}{point length(p13)/2 of p13
              -- point 11length(p13)/20 of p13}
  \fmfi{dots}{point 9length(p14)/20 of p14
              -- point 11length(p14)/20 of p14}
  \fmfi{dots}{point length(p15)/2 of p15
              -- point 11length(p15)/20 of p15}
  \end{fmfgraph*}   
\end{fmffile} 
\end{center}

The free scalar propagator is 
\begin{equation}
    \langle \phi^{I}_{a}(x)\phi_{bJ}^{\dagger}(y) \rangle_{_{(0)}} = 
    \frac{\delta_{ab}}{(2\pi)^{2}}\frac{\delta^{I}_{J}}{(x-y)^{2}} 
    \; ,
    \label{freeprop}
\end{equation}
hence at tree-level the Green function (\ref{extrcorr}) reads
\begin{eqnarray}
    &&\hspace*{-0.6cm} G_{_{(0)}}(x,x_{1},\ldots,x_{n}) = 
    \frac{c_{N}}{(2\pi)^{2\ell}}
    \frac{1}{(x-x_{1})^{2\ell_{1}}\ldots (x-x_{n})^{2\ell_{n}}}
    \sum_{{\rm perms}~\sigma}\left[ \rule{0pt}{15pt} \tr\left( 
    T^{a_{\sigma(1)}}T^{a_{\sigma(2)}}\ldots T^{a_{\sigma(\ell)}} 
    \right) \cdot \right. \nonumber \\
    &&\hspace*{-0.6cm} \left. \rule{0pt}{15pt} \cdot 
    \tr \left( T^{a_{1}}\ldots T^{a_{\ell_{1}}}\right) 
    \tr\left(T^{a_{\ell_{1}+1}}\ldots T^{a_{\ell_{1}+\ell_{2}}}\right) 
    \ldots \tr\left(T^{a_{\ell_{1}+\ldots+\ell_{n-1}+1}}\ldots T^{a_{\ell}}
    \right) \right] \; .
\end{eqnarray}

Using the ${\cal N}$=1 description introduced in the previous section,
we will momentarily show that the first order perturbative correction to 
(\ref{extrcorr}) and hence to (\ref{qextrcorr}) 
is zero. Let us preliminarily notice that if one keeps all the component 
fields in the vector supermultiplets, \ie if one does not employ the WZ 
gauge, and works in the Fermi-Feynman gauge $\alpha=1$ there is 
no first order correction to the propagators of the elementary fields. 
The off-shell self-energy corrections due to vector exchange, including a 
very peculiar $C-D$ exchange, cancel the contributions due to the three
chiral multiplets \cite{kov}. In this approach, we do not need bother 
with UV and IR problematic corrections to two-point functions of 
elementary fields. 

It is easier to first analyse the possible corrections to the 
relevant diagrams in superfield language, in which (\ref{extrcorr}) is 
obtained as the $\theta=0$ component of a correlator of (anti)chiral 
superfields. The choice of flavour indices made in (\ref{extrcorr}) is 
crucial in all subsequent computations. In particular, one can easily check 
that it prevents the insertion of (anti)chiral trilinear vertices at 
order $g^{2}$. The only relevant diagrams are thus 
obtained by the insertion of vector superfield lines into the tree 
diagram. It is very convenient to regularise the diagrams by 
point-splitting. 
According to the form of the action in components, see equations 
(\ref{sfree}) and (\ref{strilinear}), vector exchange between a 
pair of chiral lines 

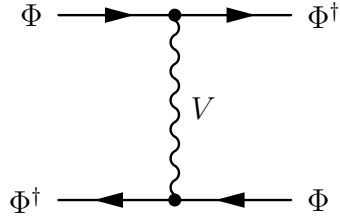
\vspace*{1cm}
\begin{center}
\begin{fmffile}{Vexc}
 \begin{fmfgraph*}(110,55) 
  \fmfleft{i1,i2} 
  \fmfright{o1,o2}
  \fmf{fermion}{o1,v1}
  \fmf{fermion}{v1,i1}
  \fmf{fermion}{i2,v2}
  \fmf{fermion}{v2,o2}
  \fmfdot{v1,v2}
  \fmffreeze
  \fmf{boson,tension=0,label=$V$}{v1,v2}
  \fmfv{la=$\Phi$,la.a=180}{i2}
  \fmfv{la=$\Phi$,la.a=0}{o1}
  \fmfv{la=$\Phi^{\dagger}$,la.a=180}{i1}
  \fmfv{la=$\Phi^{\dagger}$,la.a=0}{o2}
  \end{fmfgraph*}   
\end{fmffile} 
\end{center}

\vspace*{0.2cm}
\noindent
corresponds to three different diagrams for the lowest scalar 
components 

\vspace*{1cm}    
\noindent
\begin{equation}
\hspace*{-0.5cm} \raisebox{-1.1cm}{
\begin{fmffile}{A}
 \begin{fmfgraph*}(115,60) 
  \fmfleft{i1,i2} 
  \fmfright{o1,o2}
  \fmf{phantom,tag=1}{i1,v1}
  \fmf{phantom,tag=2}{v1,o1}
  \fmf{phantom,tag=3}{o1,v1}
  \fmf{phantom,tag=4}{v1,i1}
  \fmf{phantom,tag=5}{i2,v2}
  \fmf{phantom,tag=6}{v2,o2}
  \fmf{phantom,tag=7}{o2,v2}
  \fmf{phantom,tag=8}{v2,i2}
  \fmf{phantom,tension=0,tag=9}{v1,v2}
  \fmf{phantom,tension=0,tag=10}{v2,v1}
  \fmfdot{v1,v2}
  \fmffreeze
  \fmfipath{p[]}
  \fmfiset{p1}{vpath1(__i1,__v1)}
  \fmfiset{p2}{vpath2(__v1,__o1)}
  \fmfiset{p3}{vpath3(__o1,__v1)}
  \fmfiset{p4}{vpath4(__v1,__i1)}
  \fmfiset{p5}{vpath5(__i2,__v2)}
  \fmfiset{p6}{vpath6(__v2,__o2)}
  \fmfiset{p7}{vpath7(__o2,__v2)}
  \fmfiset{p8}{vpath8(__v2,__i2)}
  \fmfiset{p9}{vpath9(__v1,__v2)}
  \fmfiset{p10}{vpath10(__v2,__v1)}
  \fmfi{scalar}{subpath (0,length(p5)) of p5}
  \fmfi{scalar}{subpath (0,length(p6)) of p6}
  \fmfi{scalar}{subpath (0,length(p3)) of p3}
  \fmfi{scalar}{subpath (0,length(p4)) of p4}
  \fmfi{boson,l=$~A_{\mu}$,l.a=0}{subpath (0,length(p9)) of p9}
  \fmfiv{l=$\phi^{\dagger}$,l.a=-90}{point length(p4) of p4}
  \fmfiv{l=$\phi^{\dagger}$,l.a=90}{point length(p6) of p6}
  \fmfiv{l=$\phi$,l.a=90}{point length(p8) of p8}
  \fmfiv{l=$\phi$,l.a=-90}{point length(p2) of p2}
 \end{fmfgraph*}   
\end{fmffile} 
}
+ 
\raisebox{-1.1cm}{
\begin{fmffile}{B}
 \begin{fmfgraph*}(115,60) 
  \fmfleft{i1,i2} 
  \fmfright{o1,o2}
  \fmf{phantom,tag=1}{i1,v1}
  \fmf{phantom,tag=2}{v1,o1}
  \fmf{phantom,tag=3}{o1,v1}
  \fmf{phantom,tag=4}{v1,i1}
  \fmf{phantom,tag=5}{i2,v2}
  \fmf{phantom,tag=6}{v2,o2}
  \fmf{phantom,tag=7}{o2,v2}
  \fmf{phantom,tag=8}{v2,i2}
  \fmf{phantom,tension=0,tag=9}{v1,v2}
  \fmf{phantom,tension=0,tag=10}{v2,v1}
  \fmfdot{v1,v2}
  \fmffreeze
  \fmfipath{p[]}
  \fmfiset{p1}{vpath1(__i1,__v1)}
  \fmfiset{p2}{vpath2(__v1,__o1)}
  \fmfiset{p3}{vpath3(__o1,__v1)}
  \fmfiset{p4}{vpath4(__v1,__i1)}
  \fmfiset{p5}{vpath5(__i2,__v2)}
  \fmfiset{p6}{vpath6(__v2,__o2)}
  \fmfiset{p7}{vpath7(__o2,__v2)}
  \fmfiset{p8}{vpath8(__v2,__i2)}
  \fmfiset{p9}{vpath9(__v1,__v2)}
  \fmfiset{p10}{vpath10(__v2,__v1)}
  \fmfi{scalar}{subpath (0,length(p5)) of p5}
  \fmfi{scalar}{subpath (0,length(p6)) of p6}
  \fmfi{scalar}{subpath (0,length(p3)) of p3}
  \fmfi{scalar}{subpath (0,length(p4)) of p4}
  \fmfi{dashes,l=$~D$,l.a=0}{subpath (0,length(p9)) of p9}
  \fmfiv{l=$\phi^{\dagger}$,l.a=-90}{point length(p4) of p4}
  \fmfiv{l=$\phi^{\dagger}$,l.a=90}{point length(p6) of p6}
  \fmfiv{l=$\phi$,l.a=90}{point length(p8) of p8}
  \fmfiv{l=$\phi$,l.a=-90}{point length(p2) of p2}
 \end{fmfgraph*}   
\end{fmffile} 
}
+ 
\raisebox{-1.1cm}{
\begin{fmffile}{C}
 \begin{fmfgraph*}(115,60) 
  \fmfleft{i1,i2} 
  \fmfright{o1,o2}
  \fmf{phantom,tag=1}{i1,v1}
  \fmf{phantom,tag=2}{v1,o1}
  \fmf{phantom,tag=3}{o1,v1}
  \fmf{phantom,tag=4}{v1,i1}
  \fmf{phantom,tag=5}{i2,v2}
  \fmf{phantom,tag=6}{v2,o2}
  \fmf{phantom,tag=7}{o2,v2}
  \fmf{phantom,tag=8}{v2,i2}
  \fmf{phantom,tension=0,tag=9}{v1,v2}
  \fmf{phantom,tension=0,tag=10}{v2,v1}
  \fmfdot{v1,v2}
  \fmffreeze
  \fmfipath{p[]}
  \fmfiset{p1}{vpath1(__i1,__v1)}
  \fmfiset{p2}{vpath2(__v1,__o1)}
  \fmfiset{p3}{vpath3(__o1,__v1)}
  \fmfiset{p4}{vpath4(__v1,__i1)}
  \fmfiset{p5}{vpath5(__i2,__v2)}
  \fmfiset{p6}{vpath6(__v2,__o2)}
  \fmfiset{p7}{vpath7(__o2,__v2)}
  \fmfiset{p8}{vpath8(__v2,__i2)}
  \fmfiset{p9}{vpath9(__v1,__v2)}
  \fmfiset{p10}{vpath10(__v2,__v1)}
  \fmfi{scalar}{subpath (0,length(p5)) of p5}
  \fmfi{scalar}{subpath (0,length(p6)) of p6}
  \fmfi{scalar}{subpath (0,length(p3)) of p3}
  \fmfi{scalar}{subpath (0,length(p4)) of p4}
  \fmfi{dashes,l=$~CD$}{subpath (0,length(p9)) of p9}	       
  \fmfiv{d.sh=cross,d.si=3mm}{point length(p9)/2 of p9}
  \fmfiv{l=$\phi^{\dagger}$,l.a=-90}{point length(p4) of p4}
  \fmfiv{l=$\phi^{\dagger}$,l.a=90}{point length(p6) of p6}
  \fmfiv{l=$\phi$,l.a=90}{point length(p8) of p8}
  \fmfiv{l=$\phi$,l.a=-90}{point length(p2) of p2}
 \end{fmfgraph*}   
\end{fmffile} 
}
\label{abcsplit}
\end{equation}

\vspace*{1cm}
\noindent
In the following we will refer to the sum of these three terms as 
vector exchange unless otherwise stated. In (\ref{abcsplit}) the 
internal propagators are respectively \cite{kov} 
\begin{eqnarray}
    \langle A^{a}_{\mu}(x)A^{b}_{\nu}(y) \rangle_{_{(0)}} &=& 
    \frac{\delta_{\mu\nu}\delta^{ab}}{(2\pi)^{2}(x-y)^{2}} 
    \; , \label{Aprop} \\
    \langle C^{a}(x)D^{b}(y) \rangle_{_{(0)}} &=& -
    \frac{\delta^{ab}}{(2\pi)^{2}(x-y)^{2}} 
    \; , \label{CDprop} \\
    \langle D^{a}(x)D^{b}(y) \rangle_{_{(0)}} &=& -
    \delta^{ab}\delta(x-y) \; . 
    \label{Dprop}
\end{eqnarray}

The resulting first order corrections to the correlator 
(\ref{extrcorr}) are of the form 
\begin{equation}
    G_{_{(1)}}(x,x_{1},\ldots,x_{n}) = g^{2}\: c(n,N) \: 
    {\cal G}(x,x_{1},x_{2},\ldots,x_{n}) \; ,
    \label{generpert}
\end{equation}
where the coefficient $c(n,N)$ comes from colour contractions and 
the spatial dependence is  encoded in the function 
${\cal G}(x,x_{1},x_{2},\ldots,x_{n})$. In order to proceed,
it is convenient to distinguish two types of corrections, those in 
which there is a vector exchange within a single rainbow and those in 
which the vector lines are inserted between two different rainbows. 
Contributions of the first kind will be denoted by $G^{(A)}_{_{(1)}}$ 
and the others by $G^{(B)}_{_{(1)}}$.

Corrections of the first kind are exactly those that appear 
in the two-point functions of CPO's and have been argued to vanish for 
a variety of reasons \cite{dfs,ehssw,intri}. Nevertheless we feel worth 
showing their vanishing by an explicit computation. 
Each of the diagrams in $G^{(A)}_{_{(1)}}$ is zero due to the vanishing 
of the corresponding contribution to the function 
${\cal G}(x,x_{1},x_{2},\ldots,x_{n})$. More precisely, for a diagram 
with the insertion of the vector in the rainbow connecting $x$ and $x_{i}$ 

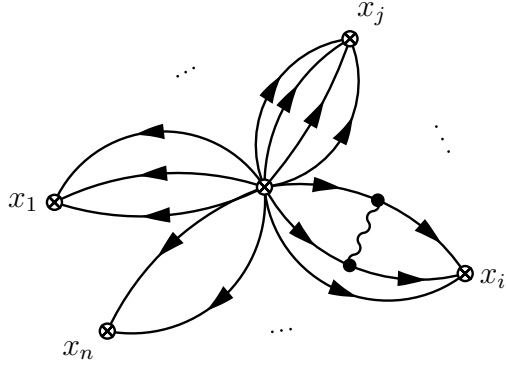
\vspace*{1cm}
\begin{center}
\begin{fmffile}{ext1a}
 \begin{fmfgraph*}(160,110) 
  \fmfsurroundn{v}{32}
  \fmf{phantom,right=0.6,tag=1}{c,v17}
  \fmf{phantom,right=0.2,tag=2}{c,v17}
  \fmf{phantom,left=0.2,tag=3}{c,v17}
  \fmf{phantom,left=0.5,tag=4}{c,v7}
  \fmf{phantom,left=0.27,tag=12}{c,v7}
  \fmf{phantom,right=0.1,tag=5}{c,v7}
  \fmf{phantom,right=0.5,tag=6}{c,v7}
  \fmf{phantom,left=0.3,tag=7}{c,v31}
  \fmf{phantom,right=0.35,tag=8}{c,v31}
  \fmf{phantom,right=0.6,tag=9}{c,v31}
  \fmf{phantom,right=0.2,tag=10}{c,v21}
  \fmf{phantom,left=0.5,tag=11}{c,v21}
  \fmf{phantom,tag=12}{v9,v25}
  \fmffreeze
  \fmf{phantom,left=0.3,tag=13}{v17,v7}
  \fmf{phantom,left=0.3,tag=14}{v7,v31}
  \fmf{phantom,left=0.15,tag=15}{v31,v21}
  \fmfipath{p[]}
  \fmfiset{p1}{vpath1(__c,__v17)}
  \fmfiset{p2}{vpath2(__c,__v17)}
  \fmfiset{p3}{vpath3(__c,__v17)}
  \fmfiset{p4}{vpath4(__c,__v7)}
  \fmfiset{p5}{vpath5(__c,__v7)}
  \fmfiset{p6}{vpath6(__c,__v7)}
  \fmfiset{p7}{vpath7(__c,__v31)}
  \fmfiset{p8}{vpath8(__c,__v31)}
  \fmfiset{p9}{vpath9(__c,__v31)}
  \fmfiset{p10}{vpath10(__c,__v21)}
  \fmfiset{p11}{vpath11(__c,__v21)}
  \fmfiset{p12}{vpath12(__c,__v7)}
  \fmfiset{p13}{vpath13(__v17,__v7)}
  \fmfiset{p14}{vpath14(__v7,__v31)}
  \fmfiset{p15}{vpath15(__v31,__v21)}
  \fmfi{fermion}{subpath (0,length(p1)) of p1}
  \fmfi{fermion}{subpath (0,length(p2)) of p2}
  \fmfi{fermion}{subpath (0,length(p3)) of p3}
  \fmfi{fermion}{subpath (0,length(p4)) of p4}
  \fmfi{fermion}{subpath (0,length(p5)) of p5}
  \fmfi{fermion}{subpath (0,length(p6)) of p6}
  \fmfi{fermion}{subpath (0,length(p7)/2) of p7}
  \fmfi{fermion}{subpath (length(p7)/2,length(p7)) of p7}
  \fmfi{fermion}{subpath (0,length(p8)/2) of p8}
  \fmfi{fermion}{subpath (length(p8)/2,length(p8)) of p8}
  \fmfi{fermion}{subpath (0,length(p9)) of p9}
  \fmfi{fermion}{subpath (0,length(p10)) of p10}
  \fmfi{fermion}{subpath (0,length(p11)) of p11}
  \fmfi{fermion}{subpath (0,length(p12)) of p12}
  \fmfi{boson}{point length(p7)/2 of p7 --
               point length(p8)/2 of p8}
  \fmfiv{d.sh=circle,d.fill=full,d.si=1.5mm}{point length(p7)/2 of p7}
  \fmfiv{d.sh=circle,d.fill=full,d.si=1.5mm}{point length(p8)/2 of p8}
  \fmfiv{d.sh=circle,d.fill=empty,d.si=2mm}{point length(0) of p1}
  \fmfiv{d.sh=cross,d.si=2mm}{point length(0) of p1}
  \fmfiv{d.sh=circle,d.fill=empty,d.si=2mm}{point length(p1) of p1}
  \fmfiv{d.sh=cross,d.si=2mm,la=$x_{1}$}{point length(p1) of p1}
  \fmfiv{d.sh=circle,d.fill=empty,d.si=2mm}{point length(p4) of p4}
  \fmfiv{d.sh=cross,d.si=2mm,la=$x_{j}$}{point length(p4) of p4}
  \fmfiv{d.sh=circle,d.fill=empty,d.si=2mm}{point length(p7) of p7}
  \fmfiv{d.sh=cross,d.si=2mm,la=$x_{i}$}{point length(p7) of p7}
  \fmfiv{d.sh=circle,d.fill=empty,d.si=2mm}{point length(10) of p10}
  \fmfiv{d.sh=cross,d.si=2mm,la=$x_{n}$}{point length(10) of p10}
  \fmfi{dots}{point length(p13)/2 of p13
              -- point 11length(p13)/20 of p13}
  \fmfi{dots}{point 9length(p14)/20 of p14
              -- point 11length(p14)/20 of p14}
  \fmfi{dots}{point length(p15)/2 of p15
              -- point 11length(p15)/20 of p15}
  \end{fmfgraph*}   
\end{fmffile} 
\end{center}

\noindent
one obtains  
\begin{eqnarray}
    && G^{(A)}_{_{(1)}}(x,x_{1},\ldots,x_{n}) = c^{(A)}(n,N) \: 
    \frac{g^{2}}{(2\pi)^{2\ell-8}}\prod^{n}_{\begin{array}{c} 
    \scriptstyle{j=1} \\ 
    \raisebox{5pt}{$\scriptstyle{j\neq i}$} \end{array}}
    \frac{1}{(x-x_{j})^{2\ell_{j}}} \cdot \nonumber \\
    && \cdot \frac{1}{(x-x_{i})^{2(\ell_{i}-2)}}\lim_{\begin{array}{c} 
    \scriptstyle{u\to x} \\ 
    \raisebox{5pt}{$\scriptstyle{v\to x_{i}}$} \end{array}}
    \left[ - \frac{1}{2(2\pi)^{10}} (\partial_{x}+
    \partial_{u})^{2} f(x,x_{i},u,v) \right] \; ,
    \label{ga-1ord}
\end{eqnarray}
where the function $f$ is defined as 
\begin{equation}
    f(x_{1},x_{2},x_{3},x_{4}) = \int dx_{5}dx_{6} \: \left[ 
    \frac{1}{x_{15}^{2}\,x_{52}^{2}\,x_{56}^{2}\,x_{36}^{2}\,x_{64}^{2}} 
    \right] \; ,
    \label{deff} 
\end{equation}
with $x_{ij}=x_{i}-x_{j}$.
In (\ref{ga-1ord}) a regularisation by point-splitting on the pair of 
lines interested by the interaction has been introduced. 
The limit $v\to x_{i}$ in (\ref{ga-1ord}) can be taken without subtleties, 
it gives a finite result \cite{ehssw} 
\begin{equation}
    \lim_{v \to x_{i}} \left[ (\partial_{x} + \partial_{u})^{2} 
    f(x,x_{i},u,v) \right] = - (2\pi)^{2} 
    \frac{(x-u)^{2}}{(x-x_{i})^{2}(u-x_{i})^{2}}\: g(x,x_{i},u) \; .
    \label{specialim}
\end{equation}
The function $g$ is defined as 
\begin{equation}
    g(x_{1},x_{2},x_{3}) = \frac{\pi^{2}}{(x_{12})^{2}} 
    B(\hat{r},\hat{s}) \; ,
    \label{defg}
\end{equation}
with $\hat{r}=\frac{x_{23}^{2}}{x_{12}^{2}}$, 
$\hat{s}=\frac{x_{13}^{2}}{x_{12}^{2}}$ and 
\begin{eqnarray}
    B(\hat{r},\hat{s}) & = & {1 \over \sqrt{p}} \left \{ \ln (\hat{r})
    \ln (\hat{s})  - 
    \left [\ln \left({\hat{r}+\hat{s}-1 -\sqrt{p}\over 2}\right)\right]^{2} 
    + \right. \nonumber \\ 
    && \left. -2 {\rm Li}_2 \left({2 \over 1+\hat{r}-\hat{s}+\sqrt {p}}
    \right ) - 2 {\rm Li}_2 \left({2 \over 1-\hat{r}+\hat{s}+\sqrt {p}}
    \right )\right \} \; ,  
    \label{Brsf}
\end{eqnarray}
where $p(\hat{r},\hat{s}) = 1+\hat{r}^{2}+\hat{s}^{2}-2\hat{r}-2\hat{s}-
2\hat{r}\hat{s}$. $g(x,x_{i},u)$ is only logarithmically divergent in the 
limit $u\to x$ \cite{ehssw,bkrs}, so that (\ref{specialim}) and thus 
(\ref{ga-1ord}) go to zero due to the prefactor $(x-u)^{2}$. In 
conclusion 
\begin{equation}
    G^{(A)}_{_{(1)}}(x,x_{1},\ldots,x_{n}) = 0 \; .
    \label{zeroga}
\end{equation}

For the second kind of first order perturbative corrections the 
insertion of vector lines between the $i$th and $j$th rainbows 
corresponds to the diagram 

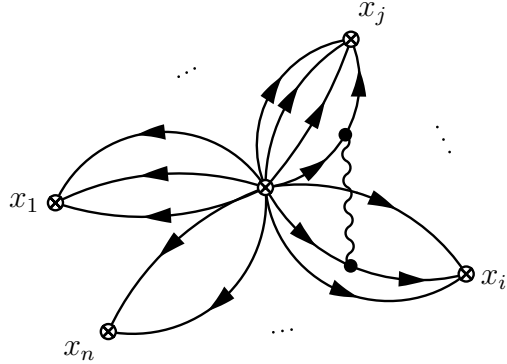
\vspace*{1cm}
\begin{center}
\begin{fmffile}{ext1b}
 \begin{fmfgraph*}(160,110) 
  \fmfsurroundn{v}{32}
  \fmf{phantom,right=0.6,tag=1}{c,v17}
  \fmf{phantom,right=0.2,tag=2}{c,v17}
  \fmf{phantom,left=0.2,tag=3}{c,v17}
  \fmf{phantom,left=0.5,tag=4}{c,v7}
  \fmf{phantom,left=0.27,tag=12}{c,v7}
  \fmf{phantom,right=0.1,tag=5}{c,v7}
  \fmf{phantom,right=0.5,tag=6}{c,v7}
  \fmf{phantom,left=0.3,tag=7}{c,v31}
  \fmf{phantom,right=0.35,tag=8}{c,v31}
  \fmf{phantom,right=0.6,tag=9}{c,v31}
  \fmf{phantom,right=0.2,tag=10}{c,v21}
  \fmf{phantom,left=0.5,tag=11}{c,v21}
  \fmf{phantom,tag=12}{v9,v25}
  \fmffreeze
  \fmf{phantom,left=0.3,tag=13}{v17,v7}
  \fmf{phantom,left=0.3,tag=14}{v7,v31}
  \fmf{phantom,left=0.15,tag=15}{v31,v21}
  \fmfipath{p[]}
  \fmfiset{p1}{vpath1(__c,__v17)}
  \fmfiset{p2}{vpath2(__c,__v17)}
  \fmfiset{p3}{vpath3(__c,__v17)}
  \fmfiset{p4}{vpath4(__c,__v7)}
  \fmfiset{p5}{vpath5(__c,__v7)}
  \fmfiset{p6}{vpath6(__c,__v7)}
  \fmfiset{p7}{vpath7(__c,__v31)}
  \fmfiset{p8}{vpath8(__c,__v31)}
  \fmfiset{p9}{vpath9(__c,__v31)}
  \fmfiset{p10}{vpath10(__c,__v21)}
  \fmfiset{p11}{vpath11(__c,__v21)}
  \fmfiset{p12}{vpath12(__c,__v7)}
  \fmfiset{p13}{vpath13(__v17,__v7)}
  \fmfiset{p14}{vpath14(__v7,__v31)}
  \fmfiset{p15}{vpath15(__v31,__v21)}
  \fmfi{fermion}{subpath (0,length(p1)) of p1}
  \fmfi{fermion}{subpath (0,length(p2)) of p2}
  \fmfi{fermion}{subpath (0,length(p3)) of p3}
  \fmfi{fermion}{subpath (0,length(p4)) of p4}
  \fmfi{fermion}{subpath (0,length(p5)) of p5}
  \fmfi{fermion}{subpath (0,length(p6)/2) of p6}
  \fmfi{fermion}{subpath (length(p6)/2,length(p6)) of p6}
  \fmfi{plain}{subpath (0,2length(p7)/5) of p7}
  \fmfi{fermion}{subpath (2length(p7)/5,length(p7)) of p7}
  \fmfi{fermion}{subpath (0,length(p8)/2) of p8}
  \fmfi{fermion}{subpath (length(p8)/2,length(p8)) of p8}
  \fmfi{fermion}{subpath (0,length(p9)) of p9}
  \fmfi{fermion}{subpath (0,length(p10)) of p10}
  \fmfi{fermion}{subpath (0,length(p11)) of p11}
  \fmfi{fermion}{subpath (0,length(p12)) of p12}
  \fmfi{boson}{point length(p6)/2 of p6 --
               point length(p8)/2 of p8}
  \fmfiv{d.sh=circle,d.fill=full,d.si=1.5mm}{point length(p6)/2 of p6}
  \fmfiv{d.sh=circle,d.fill=full,d.si=1.5mm}{point length(p8)/2 of p8}
  \fmfiv{d.sh=circle,d.fill=empty,d.si=2mm}{point length(0) of p1}
  \fmfiv{d.sh=cross,d.si=2mm}{point length(0) of p1}
  \fmfiv{d.sh=circle,d.fill=empty,d.si=2mm}{point length(p1) of p1}
  \fmfiv{d.sh=cross,d.si=2mm,la=$x_{1}$}{point length(p1) of p1}
  \fmfiv{d.sh=circle,d.fill=empty,d.si=2mm}{point length(p4) of p4}
  \fmfiv{d.sh=cross,d.si=2mm,la=$x_{j}$}{point length(p4) of p4}
  \fmfiv{d.sh=circle,d.fill=empty,d.si=2mm}{point length(p7) of p7}
  \fmfiv{d.sh=cross,d.si=2mm,la=$x_{i}$}{point length(p7) of p7}
  \fmfiv{d.sh=circle,d.fill=empty,d.si=2mm}{point length(10) of p10}
  \fmfiv{d.sh=cross,d.si=2mm,la=$x_{n}$}{point length(10) of p10}
  \fmfi{dots}{point length(p13)/2 of p13
              -- point 11length(p13)/20 of p13}
  \fmfi{dots}{point 9length(p14)/20 of p14
              -- point 11length(p14)/20 of p14}
  \fmfi{dots}{point length(p15)/2 of p15
              -- point 11length(p15)/20 of p15}
  \end{fmfgraph*}   
\end{fmffile} 
\end{center}

\noindent 
which gives 
\begin{eqnarray}
    && G^{(B)}_{_{(1)}}(x,x_{1},\ldots,x_{n}) = c^{(B)}(n,N) \: 
    \frac{g^{2}}{(2\pi)^{2\ell-8}}\prod^{n}_{\begin{array}{c} 
    \scriptstyle{k=1} \\ 
    \raisebox{5pt}{$\scriptstyle{k\neq i,j}$} \end{array}}
    \frac{1}{(x-x_{k})^{2\ell_{k}}} \cdot \nonumber \\
    && \cdot \frac{1}{(x-x_{i})^{2(\ell_{i}-1)}(x-x_{j})^{2(\ell_{j}-1)}}
    \, \lim_{u\to x} \left[ - \frac{1}{2(2\pi)^{10}} (\partial_{x}+
    \partial_{u})^{2} f(x,x_{i},u,x_{j}) \right] \; .
    \label{gb-1ord}
\end{eqnarray}
In this case the limit $u\to x$ (removal of point-splitting) is finite 
but non-vanishing, however the 
group-theory coefficient $c^{(B)}(n,N)$ turns out to be zero. This can be 
proven as follows. For compactness of notation and without loss of 
generality we consider the case in which the interaction is between the 
rainbows $i$=1 and $j$=2. One then obtains
\begin{eqnarray}
    c^{(B)}(n,N) &=& C_{N} \sum^{\ell_{1}}_{\begin{array}{c} 
    \scriptstyle{i,r=1} \\ 
    \raisebox{5pt}{$\scriptstyle{i\neq r}$} \end{array}} 
    \! \sum^{\ell_{1}+\ell_{2}}_{\begin{array}{c} 
    \scriptstyle{j,s=\ell_{1}+1} \\ 
    \raisebox{5pt}{$\scriptstyle{j\neq s}$} \end{array}}
    \! \left\{ \sum_{{\rm perms}~\sigma}\left[ 
    \rule{0pt}{15pt} \tr\left( T^{a_{\sigma(1)}} \ldots 
    T^{a_{\sigma(i)}}\ldots T^{a_{\sigma(j)}}\ldots T^{a_{\sigma(\ell)}} 
    \right) \cdot \right. \right. \nonumber \\
    && \; \cdot \:
    \tr \left( T^{a_{1}}\ldots T^{a_{r}}\ldots T^{a_{\ell_{1}}}\right) 
    \tr\left(T^{a_{\ell_{1}+1}}\ldots T^{a_{s}}\ldots 
    T^{a_{\ell_{1}+\ell_{2}}}\right) \ldots \cdot \nonumber \\ 
    && \left. \left.\rule{0pt}{15pt} \cdot \tr\left(T^{a_{\ell_{1}+
    \ldots+\ell_{n-1}+1}}\ldots T^{a_{\ell}}\right) \: 
    f_{a_{i}ba_{r}}f_{a_{j}ba_{s}} \right] \right\} = \nonumber \\
    && = - C_{N} \, \sum_{i=1}^{\ell_{1}} \,  
    \sum_{j=\ell_{1}+1}^{\ell_{1}+\ell_{2}} \, 
    \left\{ \sum_{{\rm perms}~\sigma}\left[ 
    \rule{0pt}{15pt} \tr\left( T^{a_{\sigma(1)}} \ldots 
    T^{a_{\sigma(i)}}\ldots T^{a_{\sigma(j)}}\ldots T^{a_{\sigma(\ell)}} 
    \right) \cdot \right. \right. \nonumber \\
    && \; \cdot \:  
    \tr \left( T^{a_{1}}\ldots [T^{a_{i}},T^{b}]\ldots T^{a_{\ell_{1}}}
    \right) \tr\left(T^{a_{\ell_{1}+1}}\ldots [T^{a_{j}},T^{b}]\ldots 
    T^{a_{\ell_{1}+\ell_{2}}} \right) \ldots \cdot \nonumber \\ 
    && \left. \left. \rule{0pt}{15pt} \cdot \tr\left(T^{a_{\ell_{1}+
    \ldots+\ell_{n-1}+1}}\ldots T^{a_{\ell}}\right) \right] \right\} \; ,
    \label{cb}
\end{eqnarray}
where the minus sign in the last equality comes from the factor of $i$ 
in the definition of 
the structure constants $f_{abc}$. Using cyclicity of the trace 
and the relation 
\begin{equation}
    \tr\Big( [M,T^{1}]T^{2}T^{3} \ldots T^{n} \Big) = 
    \sum_{i=2}^{n} \tr \Big( M T^{2} \ldots[T^{1},T^{i}]\ldots 
    T^{n} \Big) \; , 
    \label{zero}
\end{equation}
valid for any matrix $M$, yields
\begin{eqnarray}
    && c^{(B)}(n,N) = - C_{N} \, \sum_{i=1}^{\ell_{1}} 
    \, \sum_{j=\ell_{1}+1}^{\ell_{1}+\ell_{2}} \, 
    \left\{ \sum_{{\rm perms}~\sigma}\left[ 
    \rule{0pt}{15pt} \tr\left( T^{a_{\sigma(1)}} \ldots 
    T^{a_{\sigma(i)}}\ldots T^{a_{\sigma(j)}}\ldots T^{a_{\sigma(\ell)}} 
    \right) \cdot \right. \right. \nonumber \\
    && \cdot \sum_{p=1}^{\ell_{1}}
    \tr\Big( T^{b} T^{a_{1}}\ldots [T^{a_{i}},T^{a_{p}}]\ldots 
    T^{a_{\ell_{1}}}\Big) \sum_{q=\ell_{1}+1}^{\ell_{1}+\ell_{2}}
    \tr\Big( T^{b}T^{a_{\ell_{1}+1}}\ldots [T^{a_{j}},T^{a_{q}}]\ldots 
    T^{a_{\ell_{1}+\ell_{2}}} \Big) \cdot \nonumber \\ 
    && \left. \left.\rule{0pt}{15pt} \ldots \cdot 
    \tr\left(T^{a_{\ell_{1}+\ldots+\ell_{n-1}+1}}\ldots 
    T^{a_{\ell}}\right) \right] \right\} = 0 \; ,
    \label{zerocb}
\end{eqnarray}
since the first factor is completely symmetric in the indices $a_{m}$  
and in particular under the exchange of the pairs 
$a_{i}$, $a_{p}$ and $a_{j}$, $a_{q}$, that enter the commutators in 
the second line. 

In conclusion extremal correlators of the form (\ref{extrcorr}) have 
zero first order perturbative corrections
\begin{equation}
    G_{_{(1)}}(x,x_{1},\ldots,x_{n}) = 0 \; .
    \label{zerog1}
\end{equation}

Notice that the same analysis can be repeated step by step in the case 
of extremal correlators involving multi-trace operators in short 
multiplets \cite{dp,fz,bkrs,skiba}. 
The vanishing of the 
corrections of the first kind, previously denoted by $G^{(A)}$, is still 
valid in this case since it does not dependent on the colour structure. 
For what concerns the second type of contributions, $G^{(B)}$, the 
argument described above still applies since it only relies on the 
symmetry of the sum over permutations, that is a consequence of Wick 
contractions.

\section{Non-perturbative non-renormalisation}
\label{non_pert}

The power of supersymmetric instanton calculus \cite{akmrv}, that has 
allowed non-perturbative tests of the AdS/SCFT correspondence 
\cite{bgkr,dhkmv} will enable us
to show that extremal correlators receive no instanton contribution
from any topological sector (labelled by $K$) and for any number of 
colours (labelled by $N$) at 
leading semiclassical order. The following argument heavily relies on 
the systematic of the gaugino zero-modes in the multi-instanton background.
An index theorem for fermions in the adjoint of $SU(N)$ tells us that 
the number of gaugino zero-modes is $2 {\cal N} K N$, where the 
factor of ${\cal N}$ takes into account the number of supersymmetries.
In the ${\cal N}$=4 case under consideration, however, only 16 of these 
gaugino zero-modes are exact zero-modes \cite{bg,bgkr,dhkmv}, \ie 
those corresponding to the 8 supersymmetry
and 8 superconformal transformations broken in the YM instanton 
background.
The remaining $8KN-16$ are lifted by the Yukawa interactions and 
appear in quadrilinear terms in the ``classical action'', \ie 
in the action obtained after expanding the fields around the  
instanton configuration. 

The gaugino zero-modes may be written as
\begin{equation}
    \lambda^A = \frac{1}{2} F_{\mu\nu} \sigma^{\mu\nu} \zeta^A + 
    \hat\lambda^A \; , 
    \label{zeromod}
\end{equation}
where $\zeta^A_{\alpha} = \eta^A_{\alpha} + 
x_{\mu} \sigma^{\mu}_{\alpha\dot{\alpha}} \bar\xi^{A\dot{\alpha}}$ 
parameterise 
the 16 exact ``geometric'' zero-modes and $\hat\lambda^A$
only involve the lifted zero-modes ($8KN-16$ of them) whose explicit 
expression will not concern us here.  
$\hat\lambda^A$ can be written in terms of the ADHM data and their 
superpartners that satisfy some complicated non-linear constraints,
collectively denoted by $\Gamma^{\rm super}_{\rm ADHM}=0$ in the 
following.
The only crucial point is that the ``true'' zero-modes neither enter 
the non-linear ``super-ADHM constraints'' $\Gamma^{\rm super}_{\rm ADHM}$ 
nor the fermion-quadrilinear term in the classical action.

Taking into account the Yukawa couplings (in ${\cal N}$=1 notation)
\begin{equation}
    {\cal L}_{\rm Y} = g \, \tr \left(\phi_I^\dagger [\lambda^0, 
    \lambda^I] + \frac{1}{2}\varepsilon_{IJK} \phi^I [\lambda^J, 
    \lambda^K] + ~{\rm h.c.} \right) 
\label{yuk}
\end{equation}
the equations of motion of the scalar fields in the instanton background
and in the presence of the gaugino zero-modes read
\begin{eqnarray}
    D^2\phi^I &=& g [\lambda^0, \lambda^I] \nonumber \\
    D^2 \phi^\dagger_I &=& \frac{1}{2}g \varepsilon_{IJK} 
    [\lambda^J, \lambda^K] \; ,
    \label{eqmot}
\end{eqnarray}
where $D$ denotes the covariant derivative in the instanton background 
and the trilinear terms due to the variation of the scalar potential 
have been neglected being of higher order in $g$.

The scalar-field solutions induced by the presence of the fermionic 
zero-modes in (\ref{eqmot}) are
\begin{equation}
    \phi^1 = \zeta^0 F_{\mu\nu} \sigma^{\mu\nu} \zeta^1 + \hat\phi^1 \; ,
    \label{phisol} 
\end{equation}
where $\hat\phi^1$ is a bilinear in the lifted gaugino zero-modes 
of ``flavour'' 0 and 1 only, and
\begin{equation}
    \phi_1^\dagger = \zeta^2 F_{\mu\nu} \sigma^{\mu\nu} \zeta^3 + 
    \hat\phi_1^\dagger \; ,
    \label{barphisol} 
\end{equation}
where $\hat\phi^\dagger_1$ is a bilinear in the lifted gaugino zero-modes of 
``flavour'' 2 and 3 only.
Substituting the instanton-induced expressions for the scalar 
fields in the extremal correlators (\ref{extrcorr}) yields
\begin{eqnarray}
    G^{(K)}_{\rm inst}(x,x_{1},\ldots,x_{n}) &=&  
    c^{(K)}(n,N,g)\, e^{-\left(\frac{8\pi^{2}}{g^{2}}+ i\theta\right)K} 
    \int d\mu^{(K)}_{N} \delta(\Gamma^{\rm super}_{\rm ADHM}) 
    \cdot \nonumber \\ 
    && \cdot (\zeta^0 F \sigma \zeta^1 + 
    \hat\phi^1)^{\ell_1+...+\ell_n}(x) \ldots (\zeta^2 F \sigma \zeta^3 + 
    \hat\phi_1^\dagger)^{\ell_n}(x_{n})  \; ,
    \label{zerononpert}
\end{eqnarray}
where, except for an obvious $d^{{16}}\zeta=d^{{8}}\eta d^{{8}}\bar\xi$ 
factor, the measure of integration $d\mu^{(K)}_{N}$ and the explicit form 
of the ``super''-ADHM constraints $\Gamma^{\rm super}_{\rm ADHM}$ are 
independent of $\zeta$'s and play no r\^{o}le in the following. 
Since only the first operator could possibly absorb the zero-modes of 
type 0 and 1, the integrand necessarily contains a factor of the form
$(\zeta^0 F \sigma \zeta^1)^4(\hat\phi^1)^{\ell-4}$. Then,
using the fact that $(\zeta)^3=0$ for any two-component Grassmann 
variable, 
\begin{equation}
    G_{\rm inst}(x,x_{1},\ldots,x_{n}) = 0 
    \label{zeroginst}
\end{equation}
immediately follows for any extremal correlator. In fact in the case 
of three- and higher-point functions extremality requires $\ell \geq 
4$. For two-point functions of CPO's, that are always extremal, 
the values $\ell=2,3$ are allowed. The 
corresponding correlators have vanishing instanton correction at 
lowest order because the 16 geometric zero-modes cannot be 
saturated in these cases. 

This proves the non-perturbative non-renormalisation of extremal 
correlators of CPO's with any number $n$ of insertion points and for all 
$N$ and $K$ to leading order in the semiclassical expansion around the 
instanton background. We will not dwell into the study of perturbative 
corrections to the above instanton result, but we expect them to vanish 
as much as perturbative corrections around the trivial vacuum configuration 
vanish. The extension to any gauge group, though technically involved, 
should be straightforward.

\section{Discussion}

The results presented in this letter lend firm support to a new prediction of 
the AdS/SCFT correspondence \cite{extremal}, 
well beyond the regime of validity of the supergravity approximation. 

It would be extremely useful to fully characterise the class of correlators 
that are (expected to be) tree-level exact in ${\cal N}$=4 SYM theory. 
From a preliminary analysis of CPO's correlators, it seems that this 
interesting class  could be restricted to those correlators that involve 
only one $SU(4)$ singlet. In addition to the extremal correlators that we 
have considered above, two- and three-point functions of CPO's belong to 
this class. As observed above, all two-point functions and some three-point 
functions are extremal too. Tests of the predictions arising for these cases
from the AdS/SCFT correspondence \cite{lmrs} have been performed from the SYM 
perspective both at one-loop \cite{dfs,bkrs,skiba,hsw} and at the 
non-perturbative level \cite{bkrs}. Judiciously using the ``bonus'' 
$U(1)_{B}$ symmetry \cite{intri} in 
the context of ${\cal N}$=4 analytic superspace~\cite{ehssw} and 
disposing of some potentially troublesome contact terms \cite{psw} 
lead to a demonstration of the non-renormalisation of two- and three-point
functions of CPO's~\cite{ehw}.
The identification of $U(1)_{B}$-violating nilpotent 
super-invariants that began at 
five points and above \cite{ehw} and were not listed in
\cite{hww} prevents one from generically extending the same argument to 
higher-point functions. Proving the absence of the relevant nilpotent 
super-invariants should allow one to prove the absence of 
quantum corrections to the extremal correlators of CPO's by simple 
algebraic means. A related issue is whether the knowledge of 
correlators of CPO's completely determines those of their super-descendants 
or there are other nilpotent super-invariants that can spoil this 
naive expectation.

It would be interesting to see whether the ${\cal N}$=4 analytic superspace 
analysis could also shed some light onto the 
non-renormalisation properties of protected multi-trace  
operators that satisfy more general shortening conditions~\cite{dp,fz}.
In particular, the vanishing of their anomalous dimensions is supported by 
explicit perturbative and non-perturbative computations (\eg for a dimension 
four double-trace operator in the ${\bf 84}$ 
representation of the $SU(4)$ R-symmetry) at weak coupling 
\cite{bkrs}, but it is still waiting for an AdS confirmation.

\vspace*{0.5cm}

{\large{\bf Acknowledgements}}

\vspace*{0.2cm}
\noindent
We would like to acknowledge Dan Freedman for early collaboration on 
the project, stimulating discussions and a careful reading of the 
manuscript. We have also benefitted from discussions with Ya.S. Stanev,
E. Sokatchev, P. West and G.C. Rossi.
Preliminary results were presented by one of us (M.B.)
at the $3^{{rd}}$ Annual TMR Conference held at SISSA, Trieste, Italy,
September 20 - 24 1999. M.B. would like to thank the Aspen Center for 
Physics, where this work was initiated, for the kind hospitality.

\end{document}